\documentclass[11pt]{article}
\usepackage{moriond,epsfig}

\def\Journal#1#2#3#4{{#1} {\bf #2}, #3 (#4)}

\def\NPB{{\em Nucl. Phys.} B}
\def\PLB{{\em Phys. Lett.}  B}
\def\PRL{\em Phys. Rev. Lett.}
\def\PRD{{\em Phys. Rev.} D}
\def\ZPC{{\em Z. Phys.} C}

\def\be{\begin{equation}}
\def\ee{\end{equation}}
\def\bea{\begin{eqnarray}}
\def\eea{\end{eqnarray}}

\def\gp{\ensuremath{\gamma^{\ast} p}}
\def\wgp{\ensuremath{W_{\gamma p}}}
\def\q{\ensuremath{Q^{2}}}
\def\t{\ensuremath{|t|}}

\def\pts{\ensuremath{p_{t}^{\ast 2}}}
\def\jpsi{\ensuremath{J/\psi}}
\begin{document}
\vspace*{4cm}
\title{VECTOR MESONS AND pQCD AT HERA}

\author{ S.\,MOHRDIECK 
(on behalf of the H1 and ZEUS collaborations)
}

\address{Max-Planck-Institut f{\"u}r Physik, F{\"o}hringer Ring 6,
80805 M{\"u}nchen, Germany}

\maketitle\abstracts{
An overview of recent results from the H1 and ZEUS collaborations on
vector meson production in electron-proton collisions at HERA
is given. For diffractive vector meson production the energy 
dependence is discussed. The dependences on different scales,
$M_{VM}$, $Q^2$ and $|t|$, 
are investigated. The data are compared to pQCD theoretical
calculations.
In addition new results on inelastic \jpsi\ electroproduction in a
kinematical range $2 < \q < 100\, \rm GeV^2$ and $50 < \wgp < 225\, \rm
GeV$ are shown. Differential cross sections are compared to
theoretical predictions 
for color singlet and color octet
contributions as well as to calculations for color singlet
contributions alone.}

\section{Introduction}\label{intor}
In $ep$ scattering vector mesons are produced by two different
production mechanisms. Diffractive vector
meson production dominates the cross section in the elasicity region
$z=\frac{P_p\cdot P_{VM}}{P_p\cdot P_{\gamma^\ast}}\simeq 
1$, while inelastic processes contribute at lower $z$ values. Here are 
$P_p,\, P_{VM}\,\rm and  \, P_{\gamma^\ast}$ the four vectors of the
proton, the vector meson and the exchanged photon.
Other kinematic variables are the square of the
four momentum exchange at the positron vertex \q , the energy of the
\gp\ system \wgp\ and, especially in diffraction, the four momentum
exchange at the proton vertex \t . 
Two different kinematical regions are distinguished:
The range of quasi-real photon exchange ($\q\simeq 0$) is called
photoproduction, while electroproduction covers the region of higher
photon virtualities ($\q > 1\,\rm GeV^2$).
\\[0.3cm]
{\it 1.1 Diffractive vector meson production}
\\[0.1cm]
In the case of diffraction a photon
emitted from the incoming positron\,\footnote{During the 1998  and
part of 1999 data taking period HERA also ran with electrons}
diffracts off the proton by a colorless exchange producing an outgoing
vector meson. Two subprocesses are distinguished: 
the elastic production in which the proton stays intact
and proton dissociation in which the proton breaks up.
In the absence of a hard scale the colorless exchange can be modeled by a
soft pomeron trajectory using the Regge approach. In this case the cross
section is predicted to rise slowly with $\wgp$.
In the presence of a hard scale vector meson production 
can be calculated in the framework of perturbative QCD. In
this case the 
colorless exchange is modeled by a pair of gluons or gluon ladder
(with the quantum numbers of the vacuum).
Therefore the cross section is proportional to the square of 
the proton gluon density and a steep rise with increasing \wgp\ is predicted.
\\[0.3cm]
{\it 1.2 Inelastic vector meson production}
\\[0.1cm]
Inelastic vector meson production is dominated by the process of
photon gluon fusion where a photon emitted by the incoming positron and a
gluon as parton of the proton produce a $q\bar{q}$ pair. This $q\bar{q}$
pair can either be produced in a color singlet state by emission of a
hard gluon or in a color octet state. The color singlet state can
directly evolve into a real vector meson, while  in the case of color
octet states additional emission of soft gluons is necessary.\\
\noindent
Here the data for \jpsi\ production are compared to two theoretical
predictions\,\cite{ref:zwirner} performed in non-relativistic QCD (NRQCD).
The color singlet model  
(CS) includes only color singlet states, while the full 
calculation (CS+CO) takes in addition color octet contributions into
account. 
\section{Results on diffractive production}\label{res_diff}
HERA offers the unique possibility
to study the dependences of diffractive processes on \q, the mass of
the vector mesons $M_{VM}^2$ and \t . 
The results presented cover a range up to $100\,\rm GeV^2$ in
\q, $20<\wgp <290\,\rm GeV$ and up to $20\,\rm GeV^2$ in \t. The
production of $\rho$, $\omega$, $\phi$, $J/\psi$ and
$\Upsilon$ mesons is studied. \\
\noindent 
\begin{figure}
\unitlength1.0cm
\hspace*{0.7cm}a)\hspace*{4.6cm}b)\hspace*{5.cm}c)\\
\hspace*{-0.5cm}\epsfig{figure=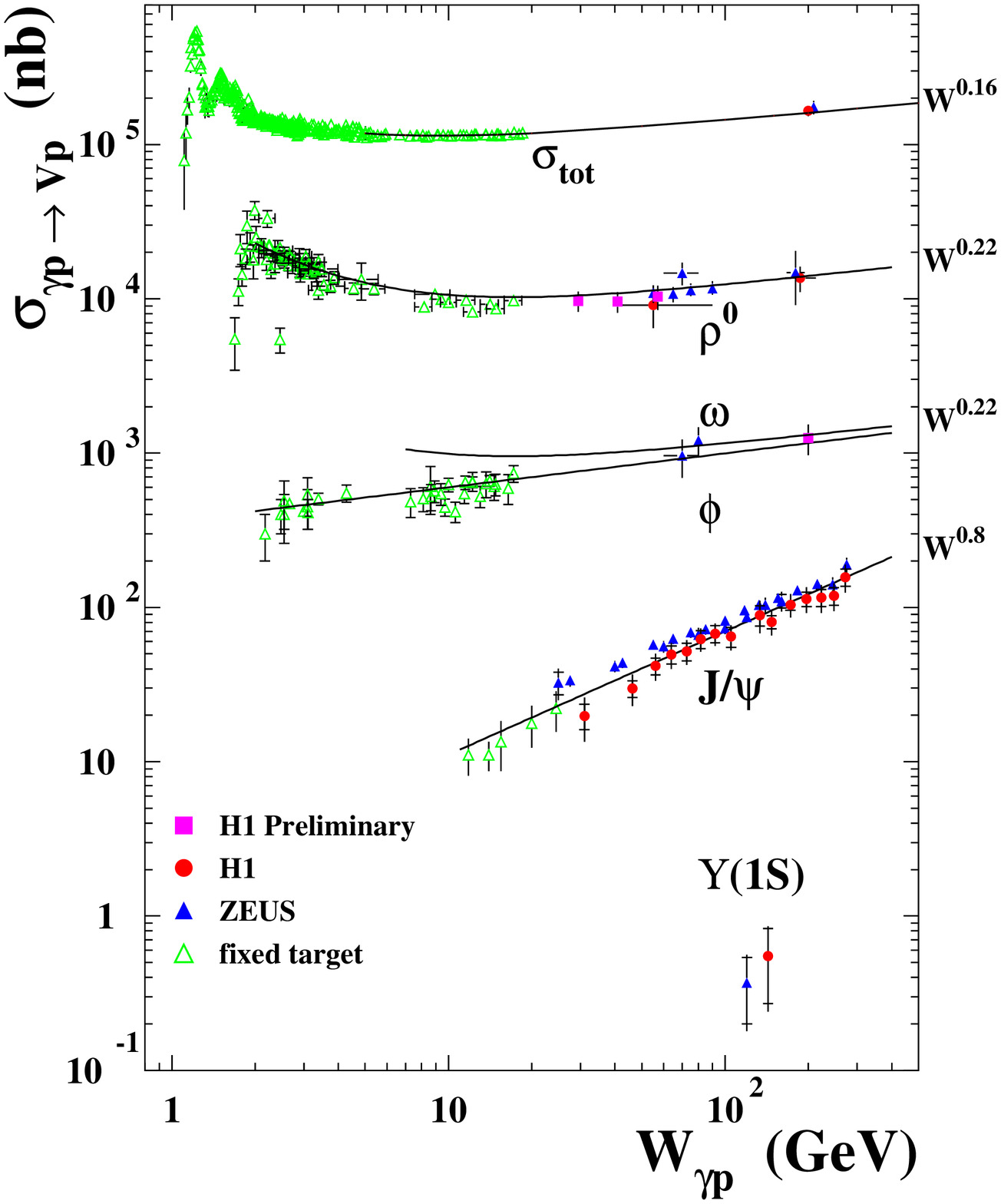,height=5.8cm}
\epsfig{figure=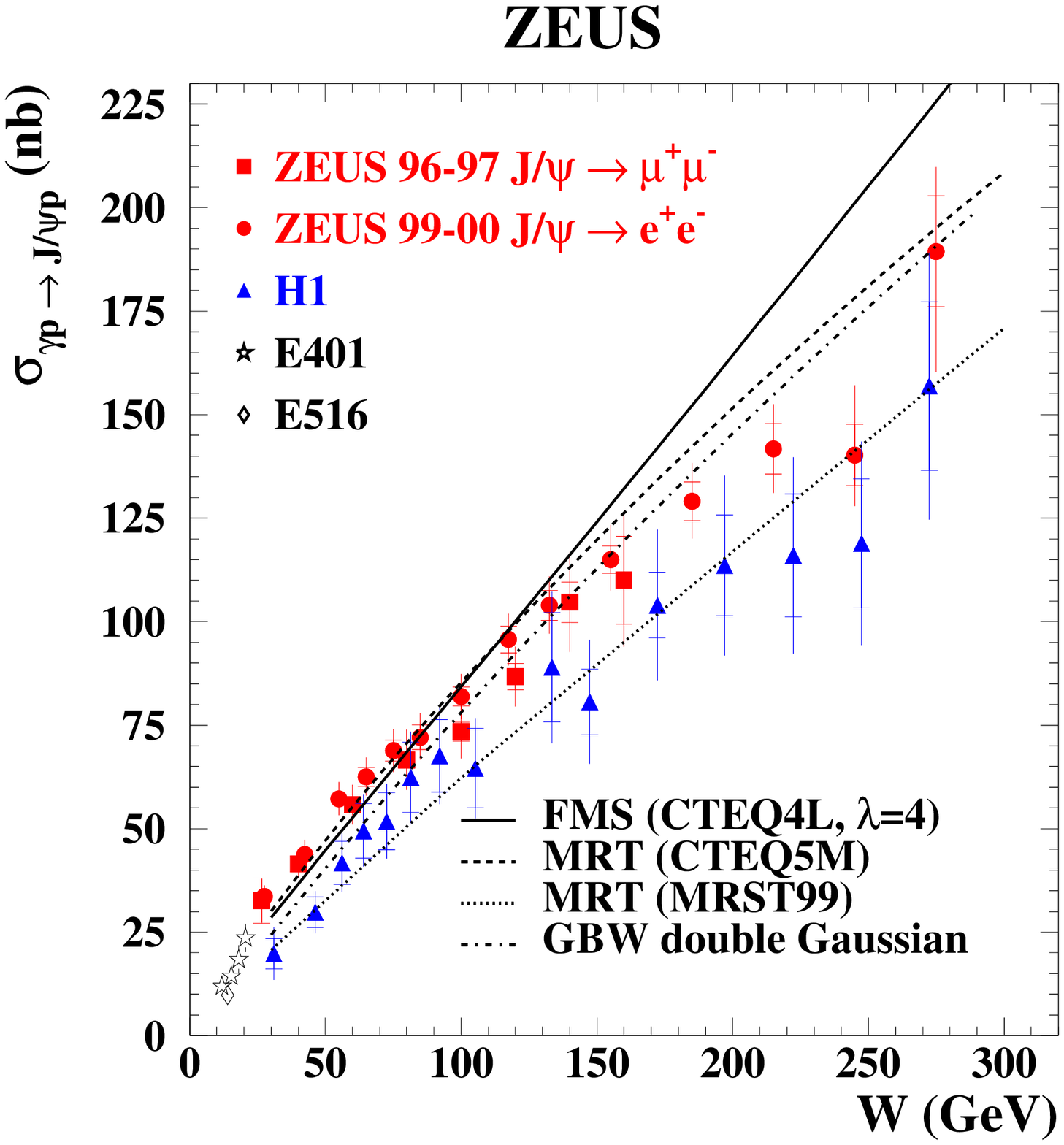,height=5.8cm}
\epsfig{figure=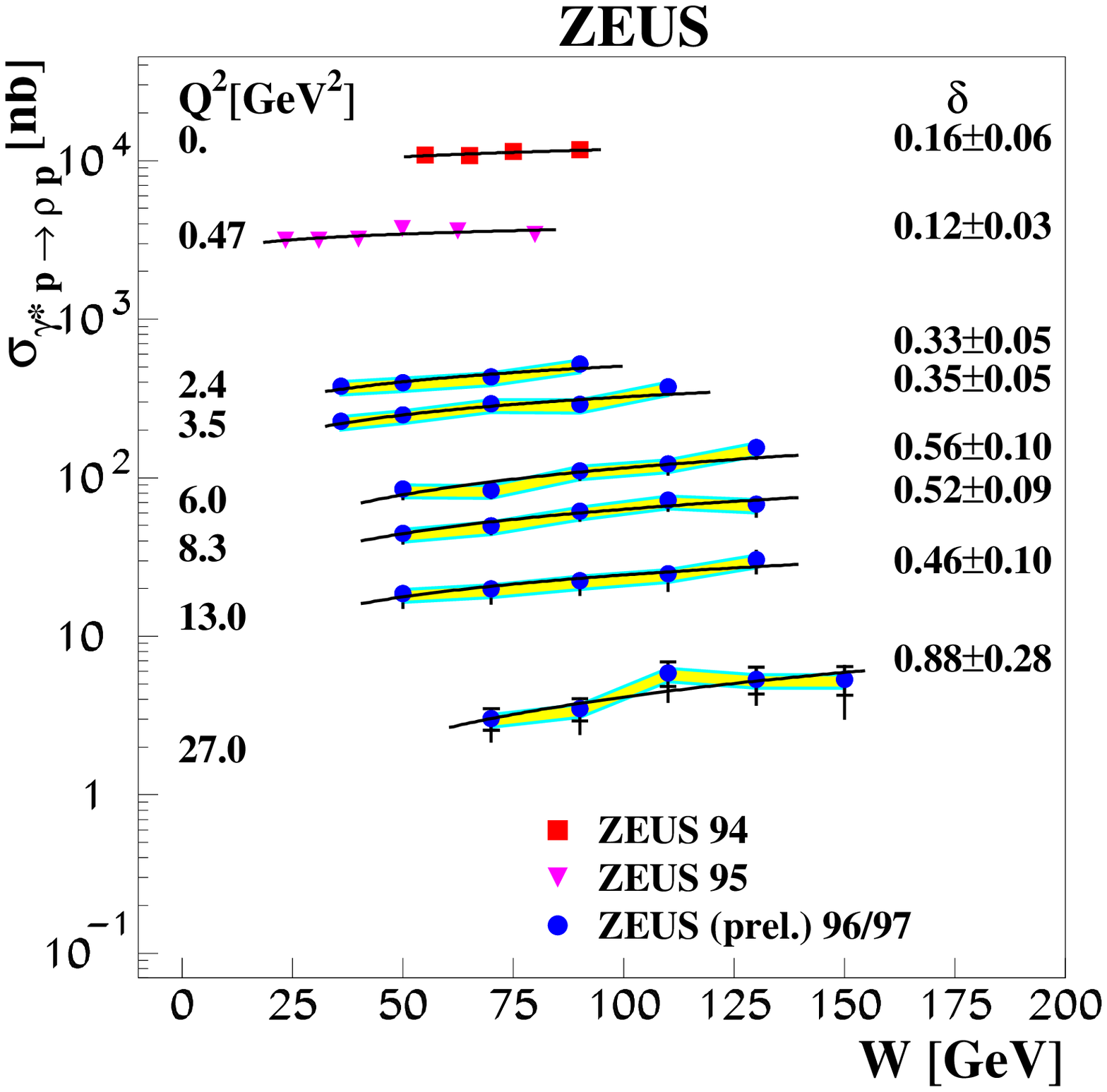,height=5.8cm}
\caption{ \label{fig:vm_w} a) \wgp\ dependence for elastic photoproduction of
different vector mesons. The lines show a behaviour of
the cross section of the form
$\wgp^{\delta}$, where $\delta$ is indicated at the right edge. 
b) \jpsi\ meson photoproduction in comparison to 
theoretical calculations in pQCD. c) \wgp\ dependence of $\rho$ meson
electroproduction at different values of \q . The lines are fits of
the form $\wgp^{\delta}$.}
\end{figure}
\noindent
The \wgp\ dependence for elastic vector meson
photoproduction is shown in figure \ref{fig:vm_w}a). 
The lines indicate a behaviour of
the cross section following a power law $\wgp^{\delta}$.
The light vector
mesons $\rho,\, \phi \,\,\rm and
\,\,\omega$ \cite{ref:lighth1,ref:lightzeus} show a 
slow \wgp\ dependence with a value for $\delta\simeq 0.22$, similar to
the value that describes
the total photoproduction cross section and well in 
agreement with predictions in the Regge approach. For the heavier vector
meson $\jpsi$
\cite{ref:heavyh1,ref:heavyzeus} a harder \wgp\ 
dependence is observed with a value $\delta\simeq 0.8$ 
as expected within pQCD 
predictions\,\cite{ref:jpsitheo}, that describe the \jpsi\ production
data very well (figure \ref{fig:vm_w}b)). 
Improved statistics are required before firm conclusions can be drawn
for the  $\Upsilon$ meson\,\cite{ref:heavyh1,ref:heavyzeus}.
Figure \ref{fig:vm_w}c) shows the developement of the \wgp\ dependence
with increasing mean values of \q\ for $\rho$ meson
production\,\cite{ref:rhodis}. Here the 
transition from a soft behaviour in photoproduction to a hard
dependence in electroproduction is seen. \\
\noindent 
\begin{figure}
\unitlength1.0cm
\hspace*{1.4cm}a)\hspace*{7.5cm}b)\\[-0.4cm]
\hspace*{0.6cm}
\epsfig{figure=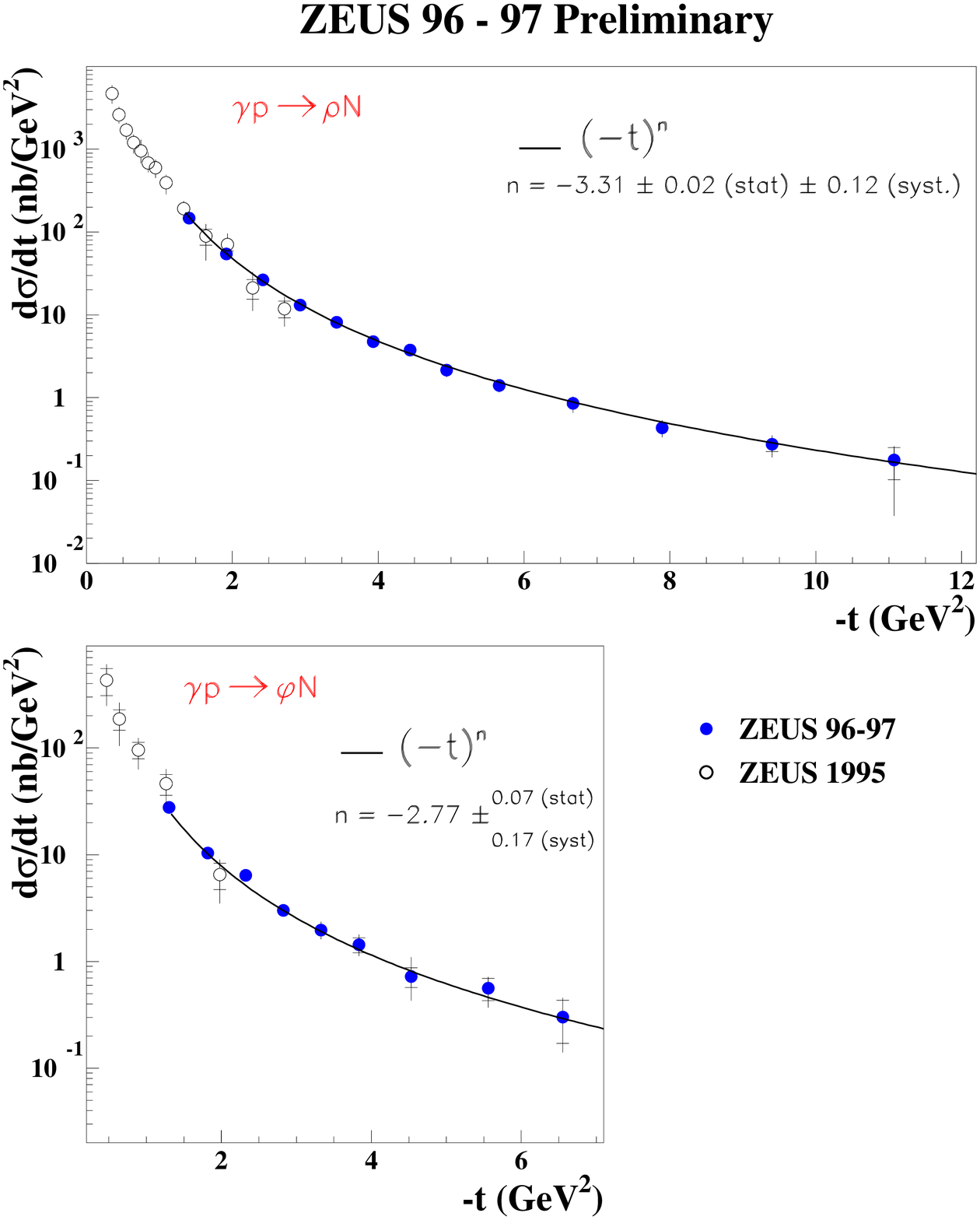,height=7.2cm}
\hfill
\epsfig{figure=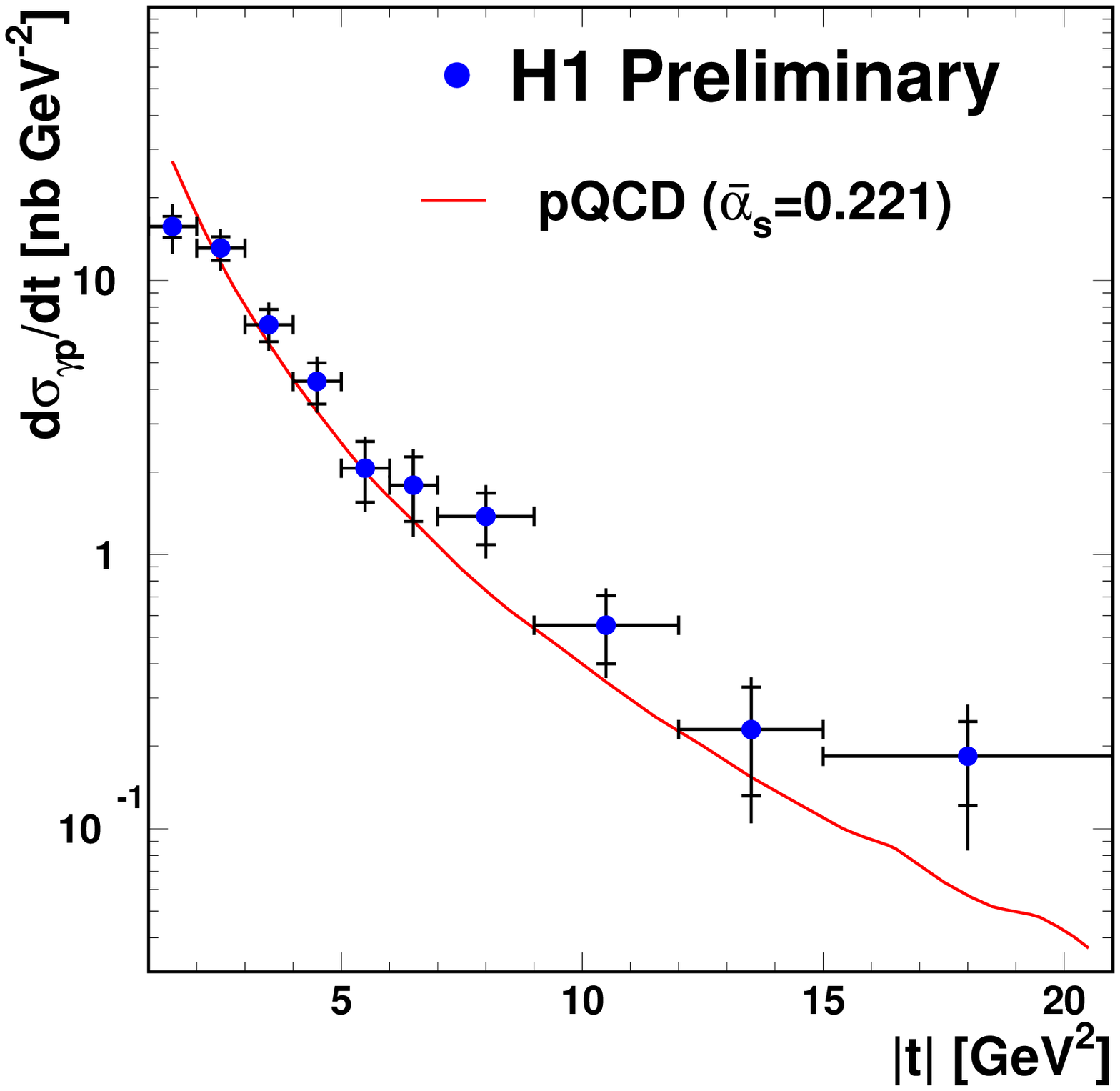,height=6.7cm}
\hspace*{0.6cm}
\caption{ \label{fig:vm_t} a) \t\ dependence in proton dissociative
photoproduction for $\rho$ and $\phi$ mesons. The lines indicate a fit
$\t^{-n}$.
b) \t\ dependence for proton dissociative \jpsi\
production in comparison to a pQCD calculation.}
\end{figure}
\noindent 
In addition to \q\ and $M_{VM}^2$ it is posdible
that \t\ may provide a hard scale. 
At high \t\ proton dissociative production dominates the
diffractive cross section. Within Regge approaches an exponential decrease
of the differential cross section in \t\ is expected. 
However, calculations\,\cite{ref:hightth} using
pQCD predict a decrease that follows a power law $\t^{-n}$.
In figure \ref{fig:vm_t}a) the differential photoproduction cross
sections as a function of \t\ for the light vector mesons $\rho$ and
$\phi$ are presented\,\cite{ref:hightzeus}. A fit to $\t^{-n}$ is 
shown  which agrees with the expectation of pQCD. 
Figure \ref{fig:vm_t}b) presents the proton
dissociative cross section of \jpsi\
photoproduction\,\cite{ref:highth1} 
compared to a pQCD calculation\,\cite{ref:forshaw}.
Good agreement is found. Both observations
indicate that \t\ provides a hard scale in diffractive vector
meson production such that pQCD can be applied at large \t .
\section{Results on inelastic production}\label{res_in}
\begin{figure}
\unitlength1.0cm
\hspace*{1.0cm}a)\hspace*{4.9cm}b)\hspace*{5.4cm}c)\\
\epsfig{figure=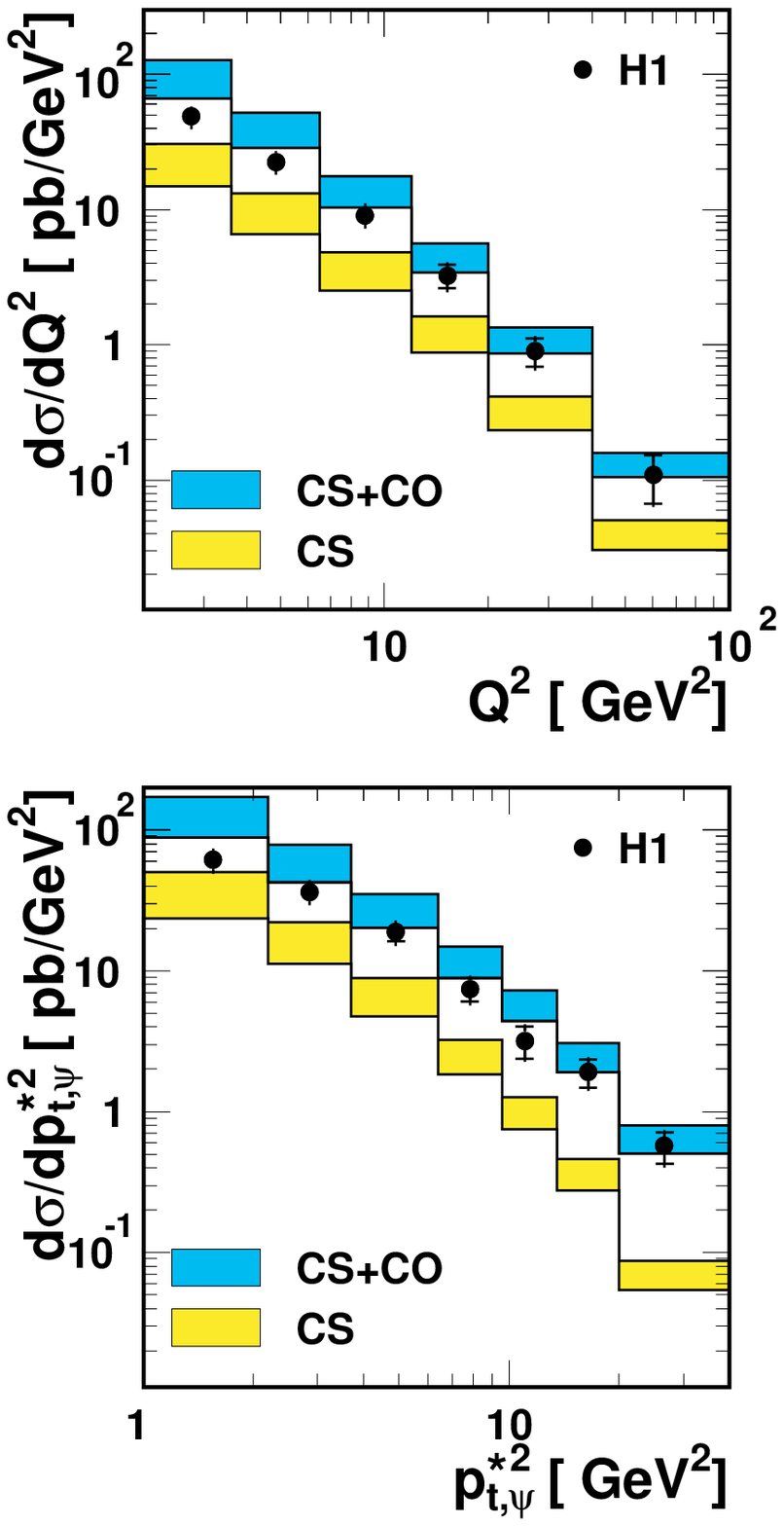,height=4.7cm,bbllx=45pt,bblly=415pt,bburx=300pt,bbury=655pt,clip=}
\epsfig{figure=fig/figlq_qfinal.2fig.eps,height=4.7cm,bbllx=45pt,bblly=165pt,bburx=300pt,bbury=405pt,clip=}
\epsfig{figure=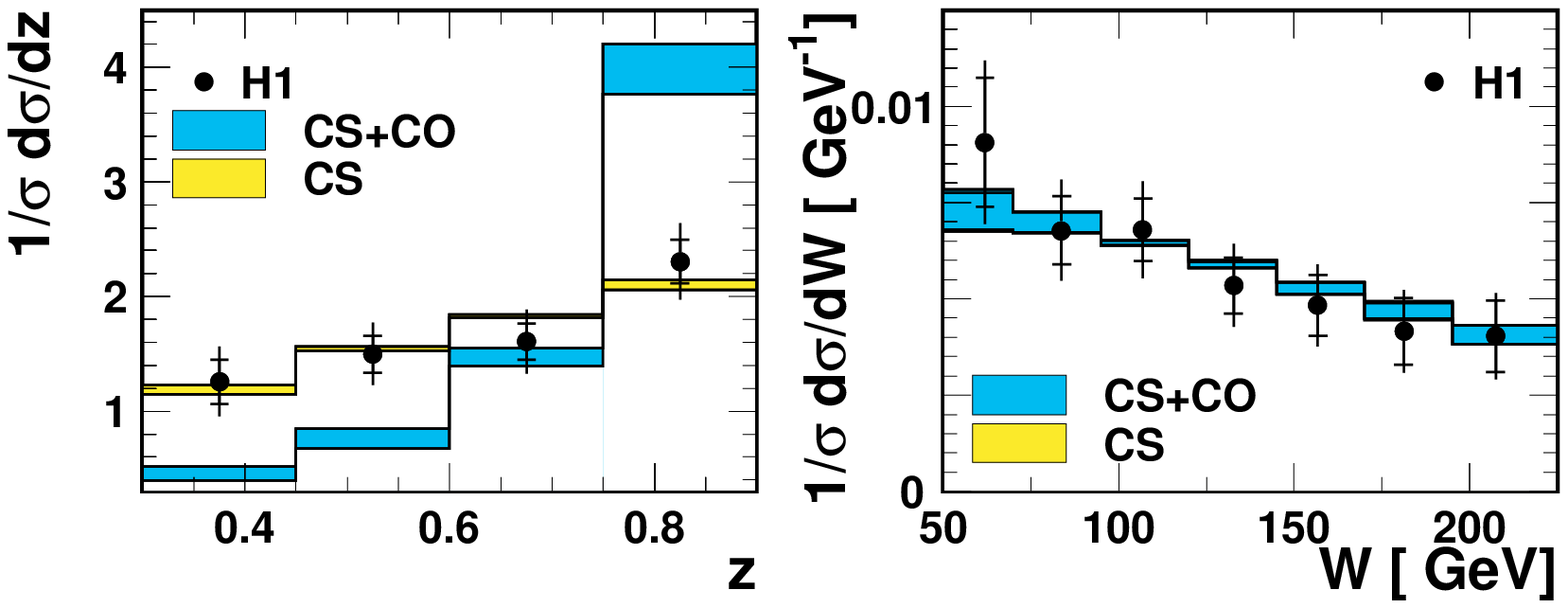,height=4.5cm,bbllx=50pt,bblly=525pt,bburx=285pt,bbury=710pt,clip=}
\caption{ \label{fig:jp_q} a) Differential $ep$ cross section in
\q\ (a), \pts\ (b) and $z$ (c) for inelastic \jpsi\
electroproduction. In c) the cross section is normalized. The data are
compared to theoretical prediction in NRQCD for the color singlet
contributions alone (CS, light grey) and for the sum of color singlet plus
color octet contributions (CS+CO, dark). }
\end{figure}

New results\,\cite{ref:paper} on inelastic \jpsi\
electroproduction in the kinematic range of $2 < \q < 100\, \rm
GeV^2$, $50 < \wgp < 225\, \rm GeV$, $0.3 < z < 0.9$ and $\pts > 1\,
\rm GeV^2$ are presented in figure \ref{fig:jp_q}.
Differential $ep$ cross sections
as a function of \q\ (a) and the squared transverse
momentum of the \jpsi\ meson in the \gp\ system \pts\ (b) are
shown as well as the normalized differential cross section in
$z$ (c). The data are compared to 
two theoretical predictions\,\cite{ref:zwirner} in leading order 
performed in the framework of NRQCD. With the light grey band the
calculation for color singlet contributions (CS) alone is shown, while the
dark band represents the full prediction including color singlet and
color octet contributions (CS+CO). The CS prediction falls below the
data by almost a factor of three and has
a too steep \pts\
dependence. Note, however, that higher order processes are
expected to contribute  significantly at high values of \pts\  as
observed in next-to-leading order calculations for CS contributions in
the photoproduction limit\,\cite{ref:kraemer}. On the other hand the
behaviour of the cross 
section in $z$ is well described. The calculation including both CS
and CO contributions shows the right depencence and magnitude of the
cross section at high \q\ and \pts , while it fails to describe the
data at high values of $z$. 
This discrepancy may be due to phase space limitations at high $z$
for the emission  of soft gluons in the transition from the color octet
$c\bar{c}$ pair to the \jpsi\ meson, which are not taken into account
in the calculation.\\
\vspace*{-0.5cm}
\section*{Conclusions}
HERA offers the unique possibility to study the transision region
from soft to hard diffraction in vector meson
production. The \wgp\ dependence of the cross section for different
vector mesons and at different values of \q\ as well as the
\t\ dependence of the cross section were presented. A clear indication
that \q , $M_{VM}^2$ and \t\ provide a hard scale is found. In the presence
of a hard scale perturbative QCD calculations describe the data well.
\\ \noindent
In addition new results on inelastic \jpsi\ electroproduction were shown and
compared to leading order theoretical predictions in NRQCD for color singlet
contributions alone and the full calculation including color singlet
and color octet contributions. The CS calculation falls below the data
by a factor of $3$, while the CO+CS prediction is in good
agreement with the data at high \q\ and \pts .
\section*{Acknowledgments}
It is a pleasure to thank the organizers for this interesting and well prepared
conference. I would like to thank my colleagues in H1 and ZEUS for providing 
the figures shown in this report.

\end{document}